\documentclass[a4paper,twoside]{article}
\newcommand{\affil}[1]{$^{\rm #1}$}
\usepackage[authoryear]{natbib}
\bibpunct{(}{)}{;}{a}{}{,}
\usepackage{graphicx}
\date{}

\title{\large\bf\flushleft Mathematical Morphology:\\
Star/Galaxy Differentiation \& Galaxy Morphology Classification}

\author{\parbox{\textwidth}{\flushleft
\vspace{-0.5cm}
{\it Jason A. Moore\affil{A,B}, Kevin A. Pimbblet\affil{A}, and Michael J.
Drinkwater\affil{A}}\\
\vspace{0.4cm}
{\small \affil{A}\,Department of Physics, University of Queensland, Brisbane QLD 4072,
Australia}\\
{\small \affil{B}\,Email: jmoore@physics.uq.edu.au}}}

\begin{document}
\twocolumn[
\begin{changemargin}{.8cm}{.5cm}
\begin{minipage}{.9\textwidth}
\vspace{-1cm}
\maketitle

\small{\bf Abstract:}

We present an application of Mathematical Morphology (MM) for the classification
of astronomical objects, both for star/galaxy differentiation and galaxy
morphology classification.  We demonstrate that, for CCD images, $99.3 \pm 3.8$\%
of galaxies can be separated from stars using MM, with $19.4 \pm 7.9$\% of the
stars being misclassified.  We demonstrate that, for photographic plate images,
the number of galaxies correctly separated from the stars can be increased using
our MM diffraction spike tool, which allows $51.0 \pm 6.0$\% of the
high-brightness galaxies that are inseparable in current techniques to be
correctly classified, with only $1.4 \pm 0.5$\% of the high-brightness stars
contaminating the population.  We demonstrate that elliptical (E) and late-type
spiral (Sc-Sd) galaxies can be classified using MM at an accuracy of
$91.4 \pm 7.8$\%.  It is a method involving less `free parameters' than current
techniques, especially automated machine learning algorithms.  The limitation of
MM galaxy morphology based on seeing and distance is also presented.  We examine
various star/galaxy differentiation and galaxy morhpology classification
techniques commonly used today, and show that the above MM techniques compare
very favourably.

\medskip{\bf Keywords:} techniques: image processing --
                        methods: data analysis --
                        methods: miscellaneous

\medskip
\medskip
\end{minipage}
\end{changemargin}
]
\small


\section{Introduction}\label{intro}

The bulk of modern astronomical observations are performed with charge-coupled
devices (CCDs) and hence are, almost by default, digital in nature.  Also, many
pre-existing sky surveys compiled with photographic plates have been digitized
using plate measuring facilities such as SuperCOSMOS \citep{ham01a, ham01b, ham01c}
and the Automatic Plate Measuring (APM) machine \citep{kib84}.  This digitization
of astronomical data has provided the opportunity for computational solutions to
many image analysis problems.  Over the past four decades, many digital
techniques have been developed for filtering, pattern recognition, neural
networks, artificial intelligence, and others.  Unsurprisingly, many of these
techniques have found applications the area of astronomical imaging.

In particular, accurately and uniformly classifying objects in astronomical
images is of great importance.  One way for this to be achieved is by performing
the classification by eye, however there are issues with this.  Firstly, eye-ball
classification is extremely subjective and it is difficult to ensure uniform
classification.  As an example, \citet{sha87} presented evidence for a possible
large-scale structure in the distribution of quasars out to redshift of about
0.5, concentrated in the direction of the cosmic microwave background dipole.
However, follow up study by \citet{dri96} concluded that there is no evidence
for such a concentration, and that the earlier result was probably biased by the
use of non-uniform image classifications.  Secondly, most astronomical
observations, particularly those for cosmological studies, seek to compile object
catalogues over large regions of the sky.  The corresponding increase in image
data (a single photographic plate can contain hundreds of thousands of detectable
objects) calls for the development of fast image processing, recognition, and
classification by automated means.  This problem of finding a robust method to
cleanly distinguish between all astronomical object types is one of the most
challenging in astronomical image analysis.

\subsection{Star/Galaxy Differentiation}\label{intro:stargalaxy}

The problem of separating the galaxy population from the stellar population
within an astronomical image is one with a variety of proposed solutions.  In
broad terms, stars and galaxies can be distinguished from each other on the basis
of their light profiles.  Stars are expected to possess a highly peak point-like
profile (the size of which is given by the point spread function of the given
image), whilst galaxies tend to be more extended in nature.  One basic technique 
is to take two parameters which describe an object in an image, plot them against
each other, and then use the line segment that optimally separates the stars from
the galaxies.  Parameters which have been commonly used (plotting against the
magnitude of the object) include: isophotal area \citep[{\textrm e.g.}][]{rei82},
peak flux \citep[{\textrm e.g.}][]{jon91}, or surface brightness
\citep[{\textrm e.g.}][]{har93}.

As an example to consider, to separate stars from galaxies in the Las Campanas
/ AAT Rich Cluster Survey (LARCS) photometric catalogue, \citet{pim01} followed
the technique of \citet{rei96}, by calculating the magnitude difference between 
4.0- and 2.0-arcsecond diameter apertures on $B$-band exposures.  Using these,
they test various other parameters of merit, including a concentration index
\citep{abr94}, full width at half maximum (FWHM), SExtractor's own estimate of
stellarity \citep{ber96}, and ellipticity.  Of these parameters, they find that
both FWHM and SExtractor both possess strong reliability and accuracy to separate
stars from galaxies.  SExtractor is very competitive as it offers not only a
rapid dissemination of results, but has the ability to reliably deblend sources
and perform star/galaxy differentiation in a highly automated fashion, thus
making the package well-suited to the reduction of large-scale galaxy survey
data.

There also exists a plethora of additional parameters which can also be helpful,
including: core magnitude, intensity weighted first moment radius, radius of
gyration, the \citet{seb79} classifier, the $r^{-2}$ `moment' of \citet{kro80},
the $Q$ classifier of \citet{lef86}, and the $\psi$ parameter of \citet{mad90}.

A more sophisticated, and less subjective, approach is to automate the
classification, using any number of the above parameters.  The classifier can be 
trained on a set of pre-classified objects by performing an automated `machine
learning' algorithm such as decision trees \citep[{\textrm e.g.}][]{wei95}, self
organizing maps \citep[{\textrm e.g.}][]{mil96}, artificial neural networks
\citep[{\textrm e.g.}][]{ode92, ode93, ber96, and00, phi02}, or fuzzy set reasoning
\citep{mah00}.  SExtractor uses a trained neural network for star/galaxy
classification.  The main problem with this technique is the large number of
`free parameters'.  Each node, the number of nodes, and the layering of the nodes
are all parameters that need to be set or tweaked within an artificial neural
network.  Additionally, the `machine learning' algorithms must be extensively
trained on pre-classified images which may introduce further biases.

\subsection{Galaxy Morphology\\Classification}\label{intro:galaxymorph}

Another challenging problem in astronomical image analysis is finding a robust
method to cleanly distinguish between all major classes of galaxies.  While most
classifications of galaxy morphology have been performed by eye using the Hubble
tuning-fork system, and its subsequent extensions
\citep[see][]{hub26, hub36, ber60, san61}, there still exist a few automated
classification techniques.

The most common technique for classifying the morphology of a galaxy is through
light decomposition.  This involves fitting a model to the surface brightness
profile of the galaxy, typically based on the Exponential, \citet{vau48, vau59}
and/or \citet{ser63, ser68} profiles.  These models effectively decompose the
galaxy profile down into bulge (de Vaucouleurs) and disk (Exponential)
components, making them a handy technique for classifying galaxies into
morphology types defined by the Hubble tuning-fork system.  As an example, GALFIT
\citep{pen02} is a galaxy and point source fitting algorithm that fits two
dimensional parameterized, axisymmetric, functions (including those mentioned
above) directly to images.

There are several issues, however, with using light decomposition.  Firstly, the
de Vaucouleurs profile does not produce the best fit for all bulges
\citep[see][]{and94, mac03}.  Secondly, even assuming a more general
Core-S\'{e}rsic model \citep[and references therein]{gra05} for the bulge, the
minimization of at least five parameters must be used to determine the `best'
solution, which is not always unique.

\citet{con03} presents an alternative solution for classifying galaxy morphology
using a three dimensional `CAS' volume: the concentration ($C$), asymmetry ($A$),
and clumpiness ($S$).  They further argue that these three parameters correlate
with important modes of galaxy evolution: gross form, major merger activity, and
star formation.  The definitions and formalism for the `CAS' parameters can be
found in Appendix~A.

\subsection{Outline}\label{intro:outline}

The plan for the remainder of the paper is as follows. In
Section~\ref{mathmorph}, the quantitative image analysis technique Mathematical
Morphology (MM) is presented, including an overview of the operators used in this
work.  In Section~\ref{stargalaxy}, the problem of star/galaxy differentiation is
investigated, including a comparison of MM with the results of current
techniques.  In Section~\ref{galaxymorph}, the problem of galaxy morphology
classification is investigated, including a comparison of MM with the results of
current techniques.  We summarize our findings in Section~\ref{summary}.


\section{Mathematical Morphology}\label{mathmorph}

MM is a branch of digital image processing and analysis originating from the work
of \citet{mat75} and \citet{ser82}, who worked on a number of problems in
mineralogy and petrology.  They laid down the foundations of MM, a new approach
to quantitative image processing.  MM has now achieved a status as a powerful
method for image processing; with applications in material science, microscopic
imaging, pattern recognition, medical imaging, and even computer vision.

The International Society for Optical Engineering (SPIE) now holds an annual
conference devoted to morphology applications, but astronomy applications have
been limited.  Some of these applications include image smoothing \citep{lea89},
removing cirrus emission \citep{app93, he96}, astronomical object extraction
\citep{can96, can97}, and quantifying simulated galaxy distributions
\citep{ued99}.

The central idea of MM is the process of examining the geometrical structure of
an image by matching it with smaller patterns at various locations in the image.
By varying the size and shape of the matching patterns, called structuring
elements\footnote{In other disciplines, the structuring element is often referred
to as a `kernel'.  In this work, however, we will stick with the original term.},
useful shape information can be obtained about the different features inside
images and their interrelations.  Although MM was originally designed to work on
only binary images, subsequent modification lead to its application to grey-scale
images.

\subsection{Structuring Elements}\label{elements}

An essential part of the MM morphological operations is the structuring element
used to probe the input image.  The structuring element is a pattern of discrete
points relative to some origin, called the active centre.  It is often chosen to
be a circle or line segment, but other choices are possible depending on the
particular application.  This however should not be viewed as a limitation, since
it actually leads to additional flexibility in algorithm design.

\begin{figure}[htbp]
\centering
\includegraphics[width=0.44\textwidth]{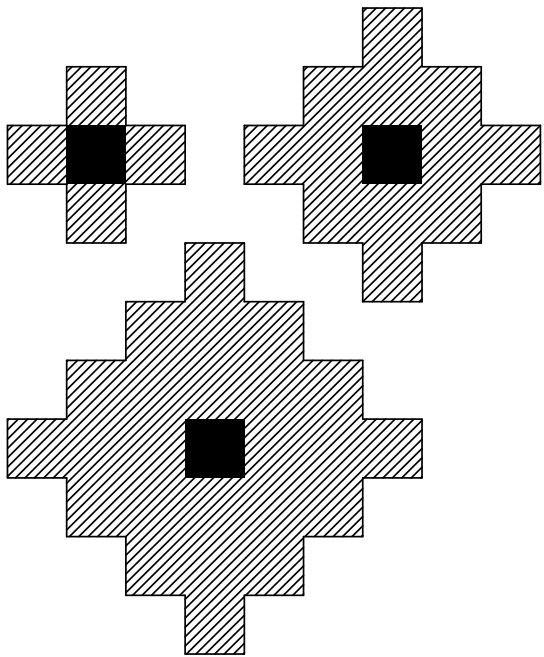}
\caption{\small{This figure illustrates an increasing series of diamond
structuring elements, created using the city-block metric.  These $3 \times 3$
(upper-left), $5 \times 5$ (upper-right) and $7 \times 7$ (bottom) structuring
elements are the first three in the series.  The shaded regions (both solid and
hatching) indicate the grid contained within the structuring elements, with the
solid fill representing the active centres.}}
\label{fig:elements}
\end{figure}

One of the most common structuring elements for best representing a circle is the
diamond shape, created using the city-block metric\footnote{A diamond can be
defined on a grid by the city-clock metric, $d(x,y) = |x| + |y|$, where $x$ and
$y$ are the grid coordinates relative to the active centre.  Positions with
$d(x,y)$ less than a defined radius are included in the shape.}.  We adopt this
as the most appropriate structuring element for the astronomical applications in
this paper, with the features of interest being approximately circular in shape.
Figure~\ref{fig:elements} contains 3 examples of the diamond structuring element,
illustrating the start of an increasing series.  The usefulness of a
monotonically increasing series of structuring elements is given in
Section~\ref{mathmorph:granulometry}.

\subsection{Operators}

In MM there are two elementary morphological operators: dilation and erosion,
from which the other operators and tools can also be formed.  Here, we give just
a brief summary of each operator, but a more detailed description of both the
operation of, and the theoretical basis for, these basic operators can be found
in the work of \citet{hei92, hei94, hei95}.

\subsubsection{Dilation Operator}

Dilation is the process of brightening, expanding, and growing regions around the
positive pixels of an image, $I$, using a structuring element, $S$.  To perform
this operation, the structuring element slides over the image.  For each position
that the structuring element is at, the maximum value contained within the
structuring element (superimposed upon the input image) is used as the pixel
value for the output image:
\begin{equation}
   (I \oplus S)_{x,y} =  \max_{(i,j) \in S} (I_{x+i,y+j})
   \label{eq:dilation}
\end{equation}

The bottom panel in Figure~\ref{fig:dilationerosion} shows a dilated image.  Note
the outward expansion of the shape that results from the operation.

\subsubsection{Erosion Operator}

Erosion is complementary to, but the opposite of dilation.  Instead of expanding,
and growing regions of the image, erosion shrinks and wipes them out.  The
process is exactly the same as for the dilation operator, save that instead of
calculating the maximum pixel value contained within the structuring element, it
calculates the minimum:
\begin{equation}
   (I \ominus S)_{x,y} =  \min_{(i,j) \in S} (I_{x+i,y+j})
   \label{eq:erosion}
\end{equation}

\begin{figure}[htbp]
\centering
\includegraphics[width=0.44\textwidth]{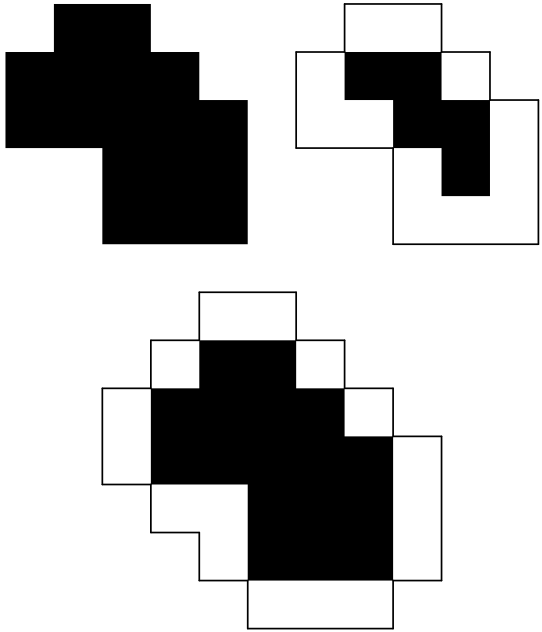}
\caption{\small{The result of performing the dilation and erosion operations on
the same shape, using a $3 \times 3$ diamond structuring element.  The upper-left
panel shows the original shape.  The bottom panel shows the dilated shape as the
outline and the original shape as the solid fill.  The upper-right panel shows
the eroded shape as the solid fill and the original shape as the outline.}}
\label{fig:dilationerosion}
\end{figure}

The upper-right panel in Figure~\ref{fig:dilationerosion} shows an eroded image.
Note the inward contraction of the shape that results from the operation.

\subsubsection{Opening Operator}

Opening is the process of separating and isolating small regions of an image from
larger ones, and simultaneously eliminating information contained on small
scales.  By sequencing an erosion followed by a dilation, the opening operator
can be formed:
\begin{equation}
   I \circ S = (I \ominus S) \oplus S
   \label{eq:open}
\end{equation}

The opening operator also has the property of being idempotent:
\begin{equation}
   (I \circ S) \circ S = I \circ S
   \label{eq:open2}
\end{equation}

This means that no more changes will occur to an image, even if the opening
operation is repeated multiple times.  The opening operator is called a
morphological filter from this property, as well as it being an increasing
morphological operator; that is, $X \subseteq Y$ implies that
$X \circ S \subseteq Y \circ S$.  The middle panel in
Figure~\ref{fig:openingclosing} shows two opened astronomical images, one of a
spiral galaxy and one of a group of elliptical galaxies.  Note the elimination of
small scale features, of the same size scale as the structuring element used,
from the image.

\subsubsection{Closing Operator}

The closing operator is the opposite of the opening operator, in a similar way as
dilation is to erosion.  Closing is the process of filling-in holes in the pixel
distribution, and joining together previously separate image features.
Mathematically, it is a dilation followed by an erosion:
\begin{equation}
   I \bullet S = (I \oplus S) \ominus S
   \label{eq:close}
\end{equation}

\begin{figure}[htbp]
\centering
\includegraphics[width=0.44\textwidth]{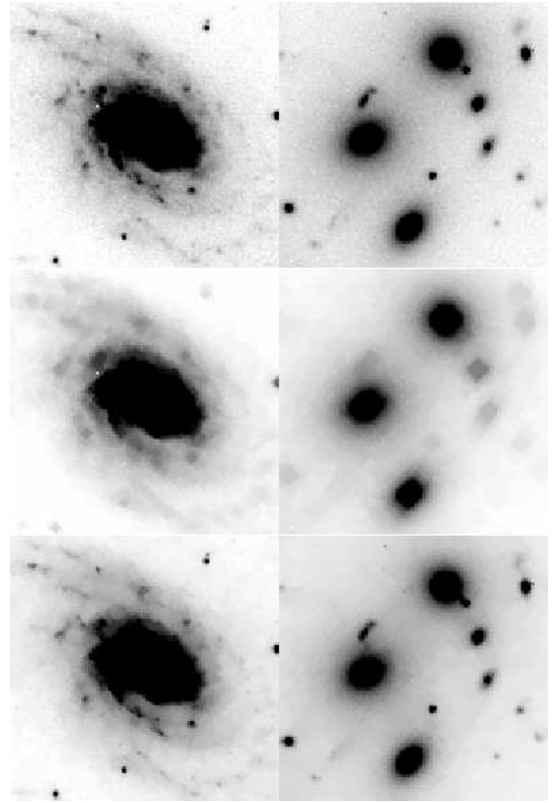}
\caption{\small{The result of performing openning and closing operations on a
spiral galaxy (left) and a group of elliptical galaxies (right).  The top row
shows the original astronomical images.  The middle row shows the opened images,
with small scale structures -- on the order of the structuring element -- being
eliminated in both.  The bottom row shows the closed images, with gaps between
features being filled in.}}
\label{fig:openingclosing}
\end{figure}

The closing operator shares the same basic features of the opening operator such
as being idempotent and working as a morphological filter.  The bottom panel in
Figure~\ref{fig:openingclosing} shows two closed astronomical images, one of a
spiral galaxy and one of a group of elliptical galaxies.

\subsubsection{Gradient Operator}

The elementary morphological operators can be combined together to detect any
edges ({\textrm i.e.} sharp gradients) around shapes in an image, as well as the
curvature of these objects.  The gradient operator is formed by combining the
dilation and erosion operators:
\begin{equation}
   \bigtriangledown_S I = (I \oplus S) - (I \ominus S)
   \label{eq:gradient}
\end{equation}

\subsection{Granulometry}\label{mathmorph:granulometry}

Granulometry uses the property of morphological operators that they can be used
to remove (or enhance) artifacts in an image of a certain size and shape,
analogous to the sieving of rocks by sequentially using sieves of progressively
larger or smaller sizes.  Granulometry consists of a sequence of closing (or
opening) operations using an increasing series of structuring elements.  By
measuring the volume under the image after successive closings, we can build a
size distribution curve:
\begin{equation}
   \Phi(\lambda) = \frac {V(\lambda) - V(0)} {V(\Lambda) - V(0)}, \lambda \geq 0
   \label{eq:sizedist2}
\end{equation}
where $\lambda$ is the parameterization of the series of closing operations,
$V (\lambda)$ is the volume of the image at each iteration, and $\Lambda$ is the
parameter associated to the largest structuring element (selected to be the one
large enough to `wipe out' the object of interest).

The size-distribution curve is monotonically increasing, so it may be considered
as a cumulative probability distribution.  The associated probability density
function is called the pattern spectrum, which is given by the following discrete
derivative:
\begin{equation}
   \Gamma(\lambda) = \Phi(\lambda + 1) - \Phi(\lambda), \lambda \geq 0
   \label{eq:pattern2}
\end{equation}

We can define a useful shape analysis attribute, the average size, which is the
expected value of the pattern spectrum:
\begin{equation}
   \bar{\lambda} = \sum_{\lambda = 0}^{\Lambda} \lambda \Gamma(\lambda)
   \label{eq:averagesize2}
\end{equation}

Another very useful attribute is the average roughness of the pattern spectrum,
which is identical to the equation for average uncertainty (entropy) in
information theory:
\begin{equation}
   \theta = - \sum_{\lambda = 0}^{\Lambda}
              \Gamma(\lambda) \log [\Gamma(\lambda)]
   \label{eq:averagerough2}
\end{equation}

\begin{figure}[htbp]
\centering
\includegraphics[width=0.44\textwidth]{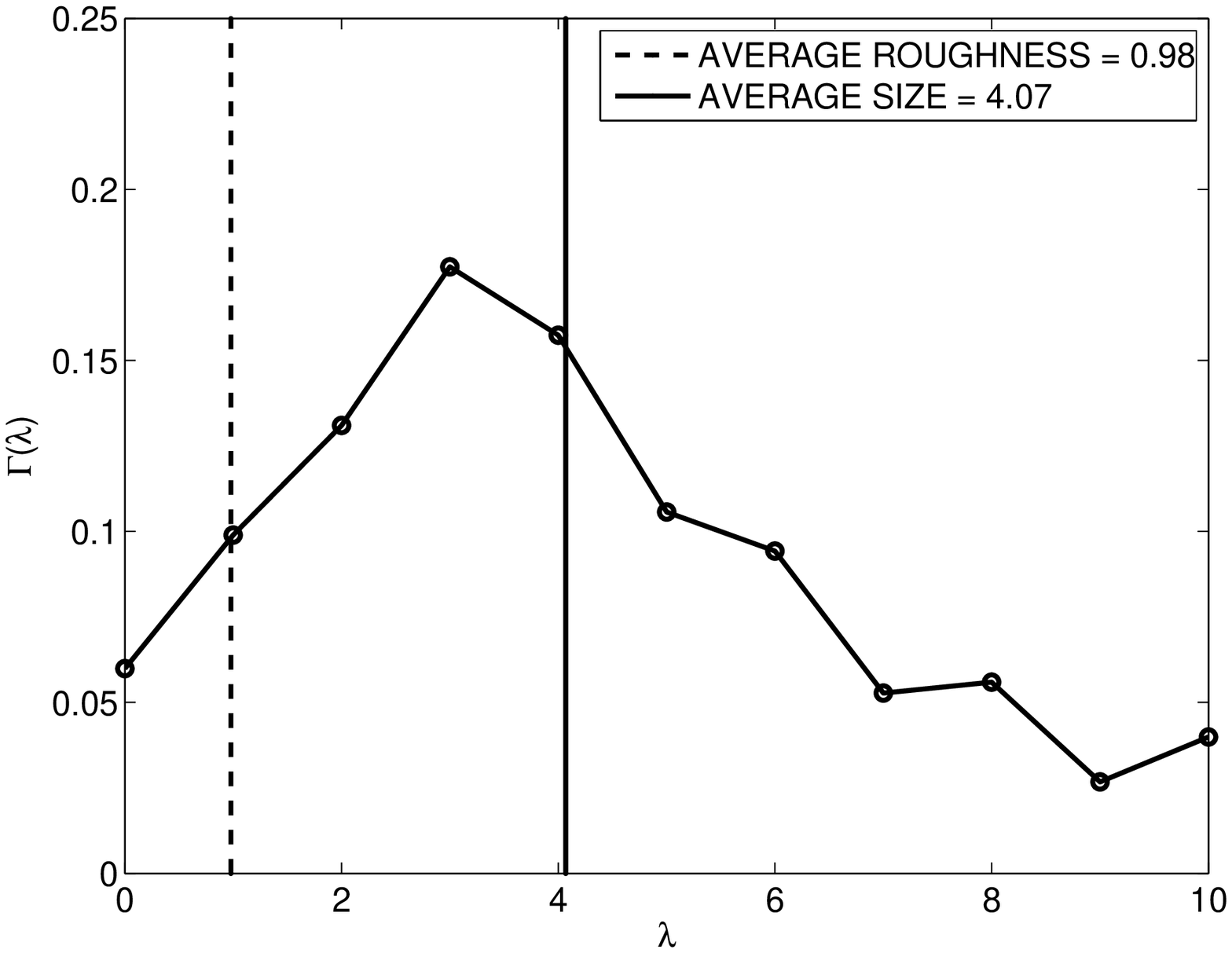}
\includegraphics[width=0.44\textwidth]{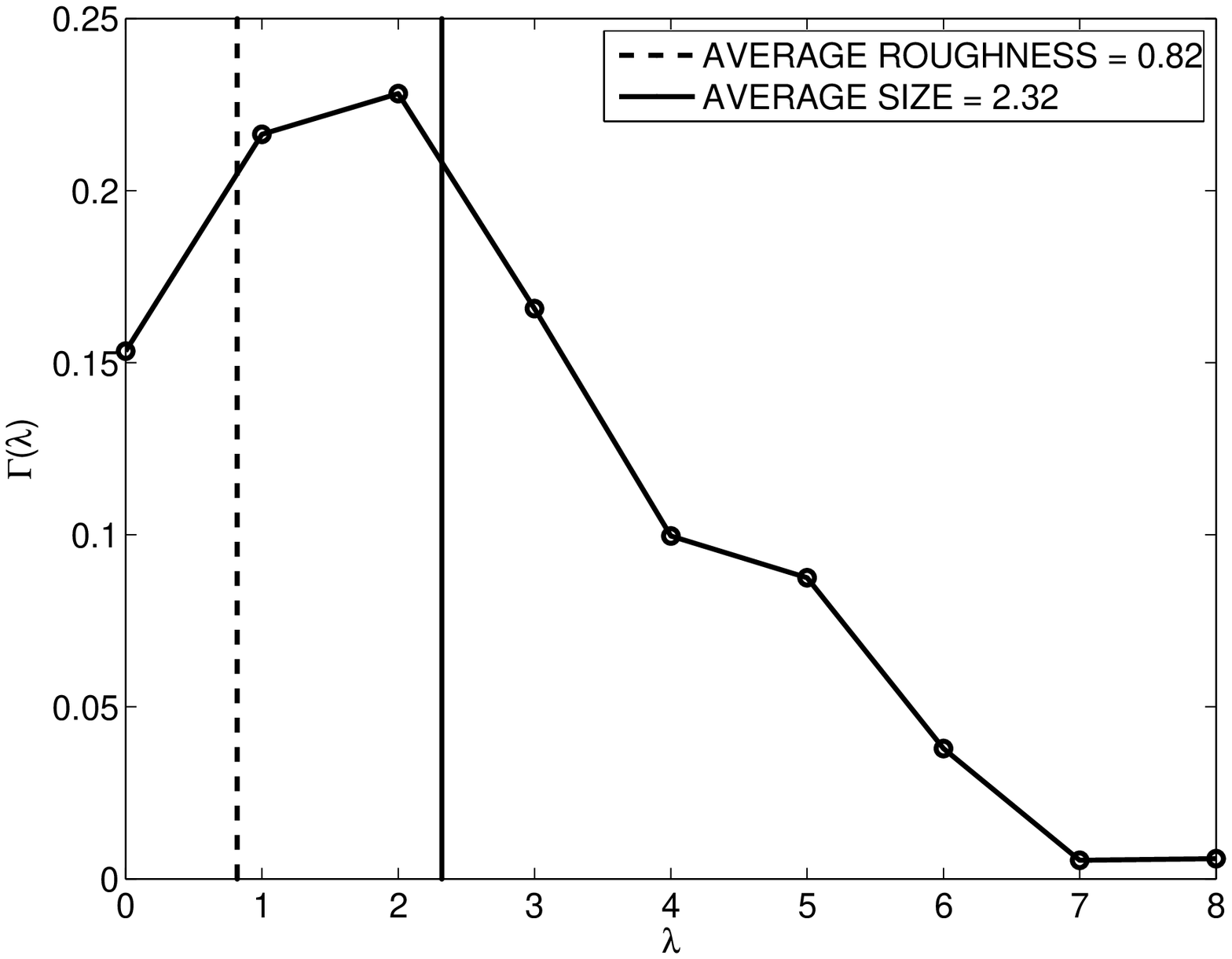}
\caption{\small{Pattern spectrum plots for CCD images of typical spiral (top) and
elliptical (bottom) galaxies.  The pattern spectrum represents the distribution
of size within an image, based on the shape of the structuring element.  The
dotted vertical lines on the plots indicate the average roughness, and the full
vertical lines represent the average size.}}
\label{fig:pattern}
\end{figure}

Figure~\ref{fig:pattern} illustrates how the pattern spectrum will look for
typical spiral nd elliptical galaxies.  Also included on the plots are the values
of average size and average roughness.

Of the MM tools derived, the average size, $\lambda$, and the average roughness,
$\theta$, are a powerful combination for characterizing shape and size in image
analysis \citep[see][]{mar89}, providing statistical information about the shape
content of the image.  In this work, we present an implementation of basic MM
operators and applications of MM in two astronomical object classification
problems: star/galaxy differentiation and galaxy morphology classification.  We
expect MM to be superior to many of the current techniques for astronomical
object classification, with the benefit of only using one `free parameter', the
structuring element, compared with at least five (ellipse fits) for the
techniques mentioned in Section~\ref{intro:stargalaxy} and
Section~\ref{intro:galaxymorph}.


\section{Star/Galaxy Differentiation}\label{stargalaxy}

To test the advantage of using MM to differentiate between stars and galaxies,
both CCD and photographic plate images are used.

CCD chips possess many qualities that make them superior to photographic plates,
including a large dynamic range, higher quantum efficiency, a higher linear
response, and relatively low noise levels.  Photographic plates can have a larger
surface area, making them perfect for capturing a wide field-of-view, however
they have have drawbacks such as the low sensitivity and non-linear response of
the plates, saturation of bright objects, high surface brightness limits, and
possible inconsistencies that exist between different plates in large surveys
(such as the SuperCOSMOS Sky Survey or the APM Sky Catalogue).

The major disadvantage in using photographic plates in astronomy is the
saturation of bright objects, such as stars or bright galaxies.  The profile of a
saturated star becomes extended, similar to that of a galaxy.  This makes it
difficult to differentiate stars from galaxies based on their light profiles.
CCD images, however, maintain most of the original light profile, giving them an
advantage in star/galaxy differentiation.  Using both astronomical image types
allows us to test the limitations of the various classification techniques.

\subsection{Classifier Performance}\label{stargalaxy:performance}

We focus on the separation of the galaxy population, which is most important in,
for example, galaxy surveys and cosmological studies.  Therefore, we rate the
performance of a star/galaxy classifier based on the following three criteria:
\begin{enumerate}
\item Completeness --- the percentage of galaxies correctly classified as such.
\item Contamination --- the percentage of stars incorrectly classified as
galaxies.
\item Correctness --- the overall percentage of objects correctly classified as
stars or galaxies.
\end{enumerate}

Rating a star/galaxy classifier by these criteria, particularly the first two,
will give us an indication of how well it extracts the galaxy population
({\textrm e.g.} for follow up spectroscopic study) without having too many stars
contaminate the sample.

In this work, a multi-layer perceptron (MLP) is used as the automated classifier
to separate the stars from the galaxies for a given star/galaxy classification
technique.  MLP \citep[see][]{hay98} is the most common neural network, and has
been applied to many complex pattern recognition problems.  It is composed of a
layering of nodes (or perceptrons) whose output is determined by a linear
combination of the previous nodes put through a nonlinear activation function.
The weighting factors within the linear combination are varied until the optimal
solution for a given set of training data is found.  Based on the number of
layers, and nodes in each layer, any function can be modeled.  In this work,
however, we have reduced the number of network nodes to the minimum to perform
basic linear classification.

\subsection{CCD Images}\label{stargalaxy:ccd}

The CCD images are taken from the data archive of \citet{sma97}, who have
produced a classified catalogue of morphological types of 1857 astronomical
objects within 10 intermediate-redshift ($0.37 < z < 0.56$) rich galaxy cluster
fields, observed using the Wide-Field Planetary Camera 2 (WFPC2) on the Hubble
Space Telescope ({\it HST}).  They classify each detectable object within the
field by eye, according to the Revised Hubble scheme --- as either a star,
spheroidal (E/S0), early-type spiral (Sa-Sb), late-type spiral (Sc-Sd), barred
spiral (SBa-SBd) or irregular (Irr/Sm) galaxy.  The reliability of these
classifications is hard to quantify, but is considered by the authors to be
accurate to within one class or better in at least 80\% of cases.  The objects
in the sample range in magnitude from 23.5 at the faint end to magnitude 15.5 at
the bright end, over the filter bands $B$ (F450W), $V$ (F555W), $R$ (F702W), and
$I$ (F814W).

We use this catalogue, both the {\it HST} images and the object segmentation
generated using SExtractor, to test the effectiveness of the various
classification techniques to differentiate between stars and galaxies in CCD
images.  We limit our results to those taken with the $R$-band filter (725
classified objects).

For these images, magnitude, aperture area, surface brightness (SB), and the MM
tools, average size and average roughness, are used to differentiate between
pre-classified stars and galaxies.  Before calculating average size and average
roughness, an MM gradient operation is performed to enhance sharp edges,
increasing the difference between star and galaxy light profiles.  To calculate
average size and average roughness, an increasing series of diamond structuring
elements is used for reasons given in Section~\ref{elements}.

\begin{figure}[htbp]
\centering
\includegraphics[width=0.44\textwidth]{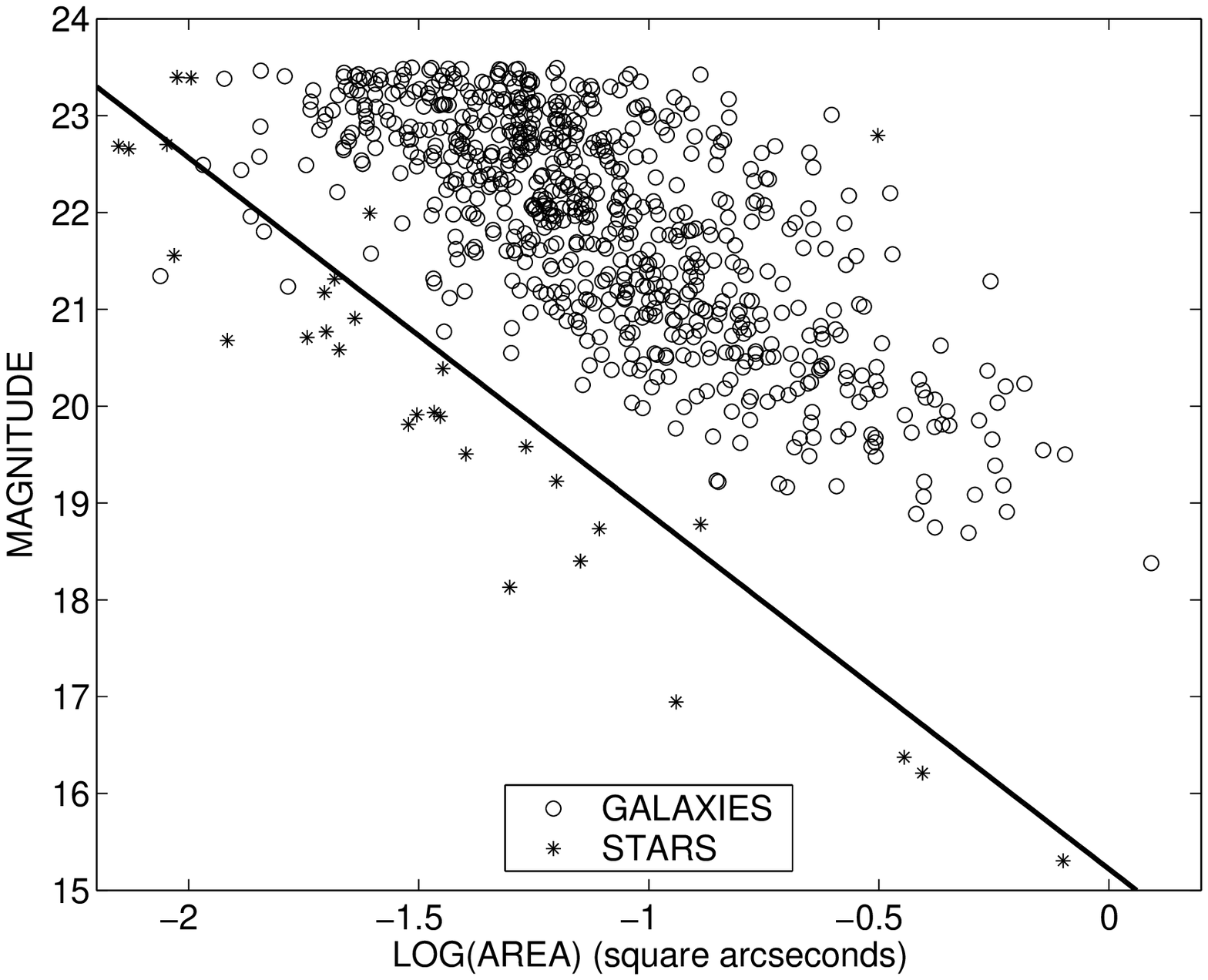}
\includegraphics[width=0.44\textwidth]{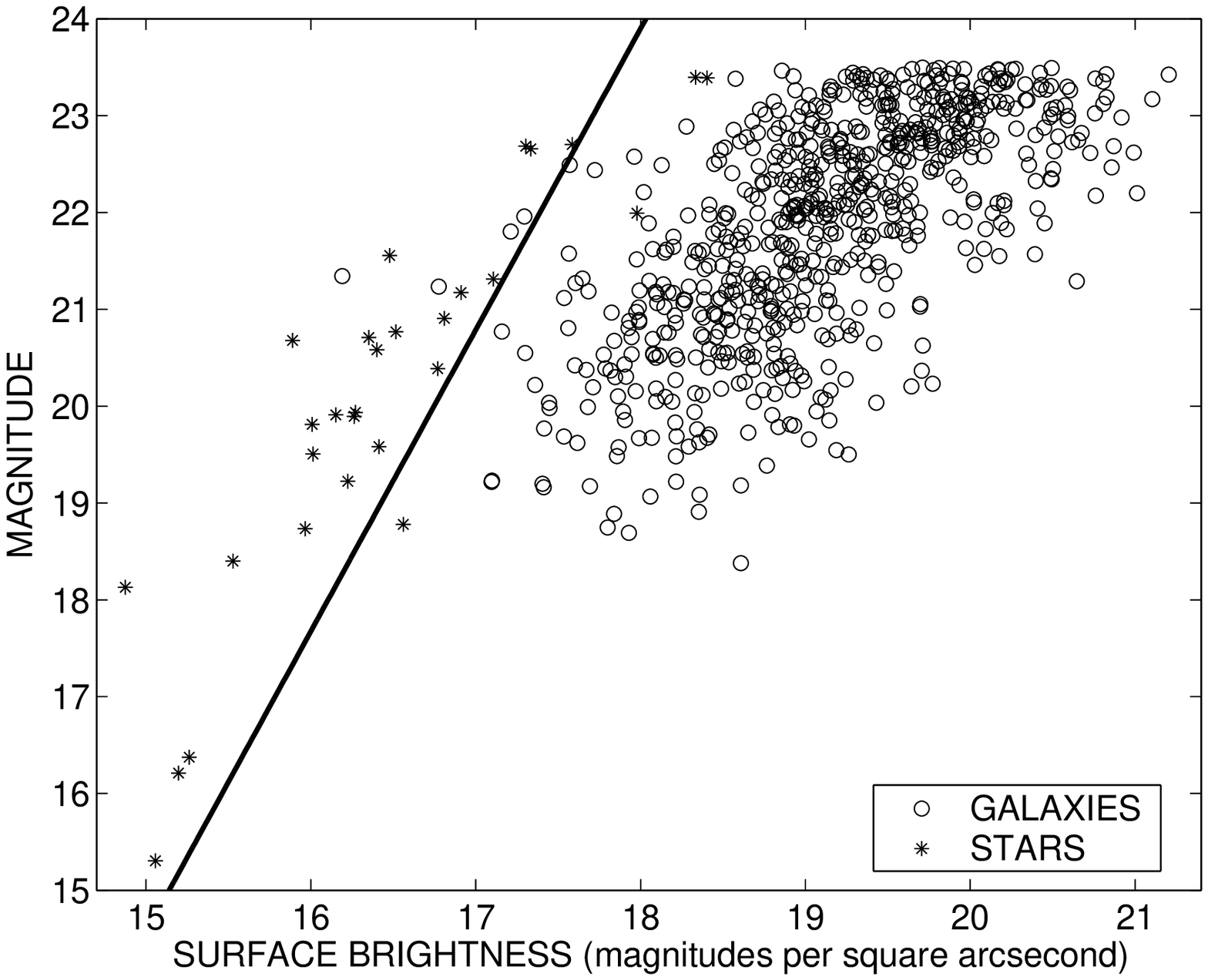}
\includegraphics[width=0.44\textwidth]{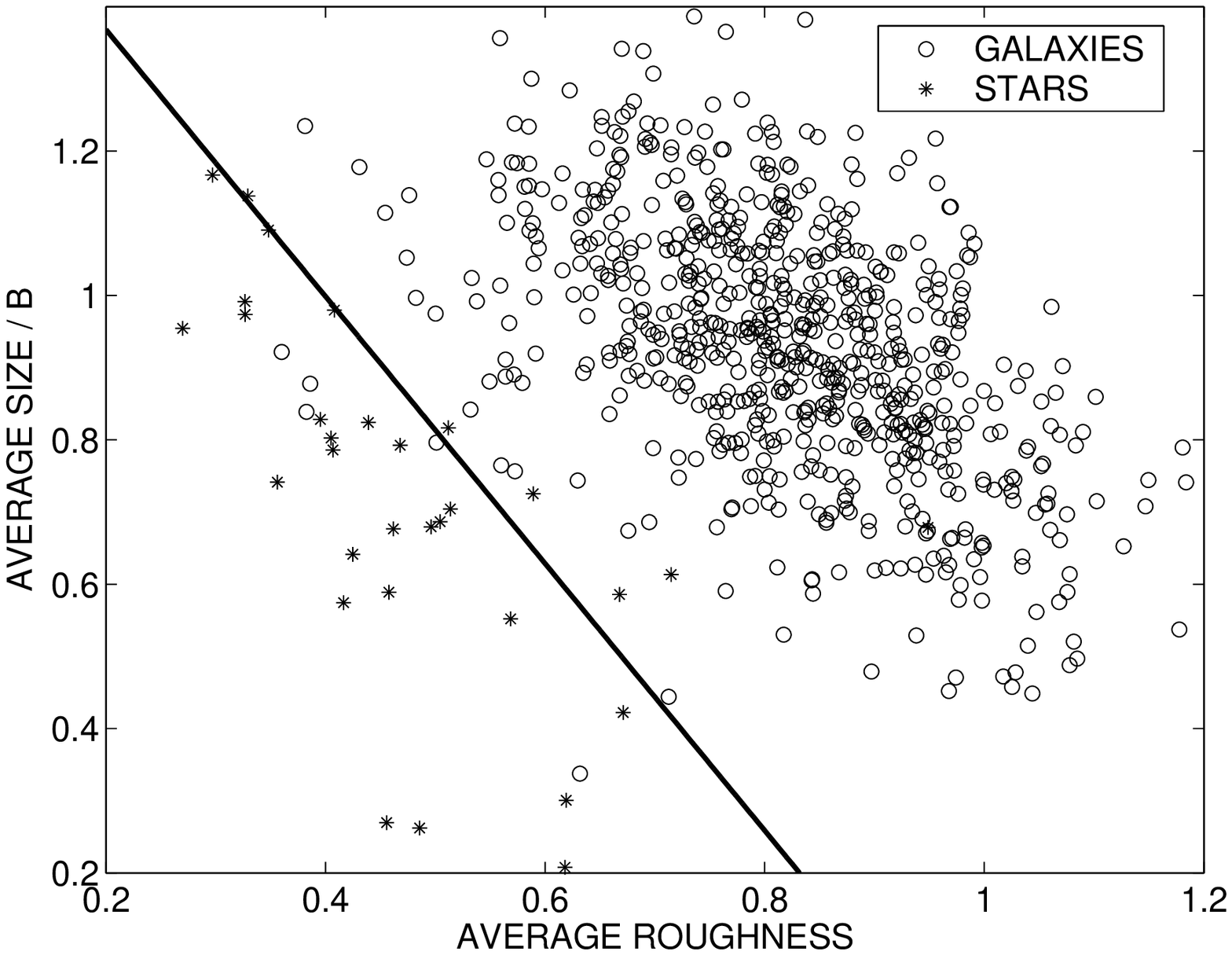}
\caption{\small{Star/galaxy classification plots for CCD image objects, between
$R$-band magnitude 23.5 and 15.5.  The upper two panels show the results of using
current techniques for the classification: the top panel showing the
classification using magnitude and aperture area, and the middle panel showing 
the classification using magnitude and surface brightness.  The bottom panel
shows our classification using two parameters from MM --- average size and
average roughness.  The classification line in each panel has been constructed
using an MLP neural network.  The classification performance of MM is comparable
to that of current techniques, in separating stars and galaxies.  These results
are summarized in Table~\ref{tab:ccd}.}}
\label{fig:ccd}
\end{figure}

Figure~\ref{fig:ccd} illustrates how magnitude, aperture area, and surface
brightness can be used to differentiate between stars and galaxies in CCD images,
and the results of using MM.  In each plot, the line segment used to separate
stars from galaxies is included --- which has been constructed using an MLP.
These results are summarized in Table~\ref{tab:ccd}.

\begin{table}[htbp]
\begin{center}
\caption{Star/Galaxy Differentiation results for CCD images$^a$.}\label{tab:ccd}
\begin{tabular}{lrrr}
\hline
Method & Comp. (\%) & Cont. (\%) & Corr. (\%) \\
\hline
Mag--Area  & $99.4 \pm 3.8$ & $16.1 \pm 7.2$ & $98.8 \pm 3.7$ \\
Mag--SB & $99.4 \pm 3.8$ & $16.1 \pm 7.2$ & $98.8 \pm 3.7$ \\
MM & $99.3 \pm 3.8$ & $19.4 \pm 7.9$ & $98.5 \pm 3.7$ \\
\hline
\end{tabular}
\medskip\\
$^a$ Using 725 objects --- 31 stars and 694 galaxies.
\end{center}
\end{table}

\subsection{Photographic Plate Images}\label{stargalaxy:photo}

\citet{doy05} use 4315 fields from the SuperCOSMOS Image Archive
\citep{ham01a, ham01b, ham01c} in their work on finding optical counterparts for
the HI Parkes All Sky Survey (HIPASS).  We use the same photographic plate
images, along with the object segmentation generated using SExtractor, to test
the effectiveness of using the various classification techniques to differentiate
between stars and galaxies in photographic plate images.  We classify each
detectable object within the field by eye in a similar way to \citet{sma97}.  The
objects in the sample range in $B$-band magnitude from 20.0 at the faint end to
magnitude 13.0 at the bright end.  For these images, magnitude, aperture area,
surface brightness (SB) and peak flux (PF) are used to differentiate between the
pre-classified stars and galaxies.

\begin{figure}[htbp]
\centering
\includegraphics[width=0.44\textwidth]{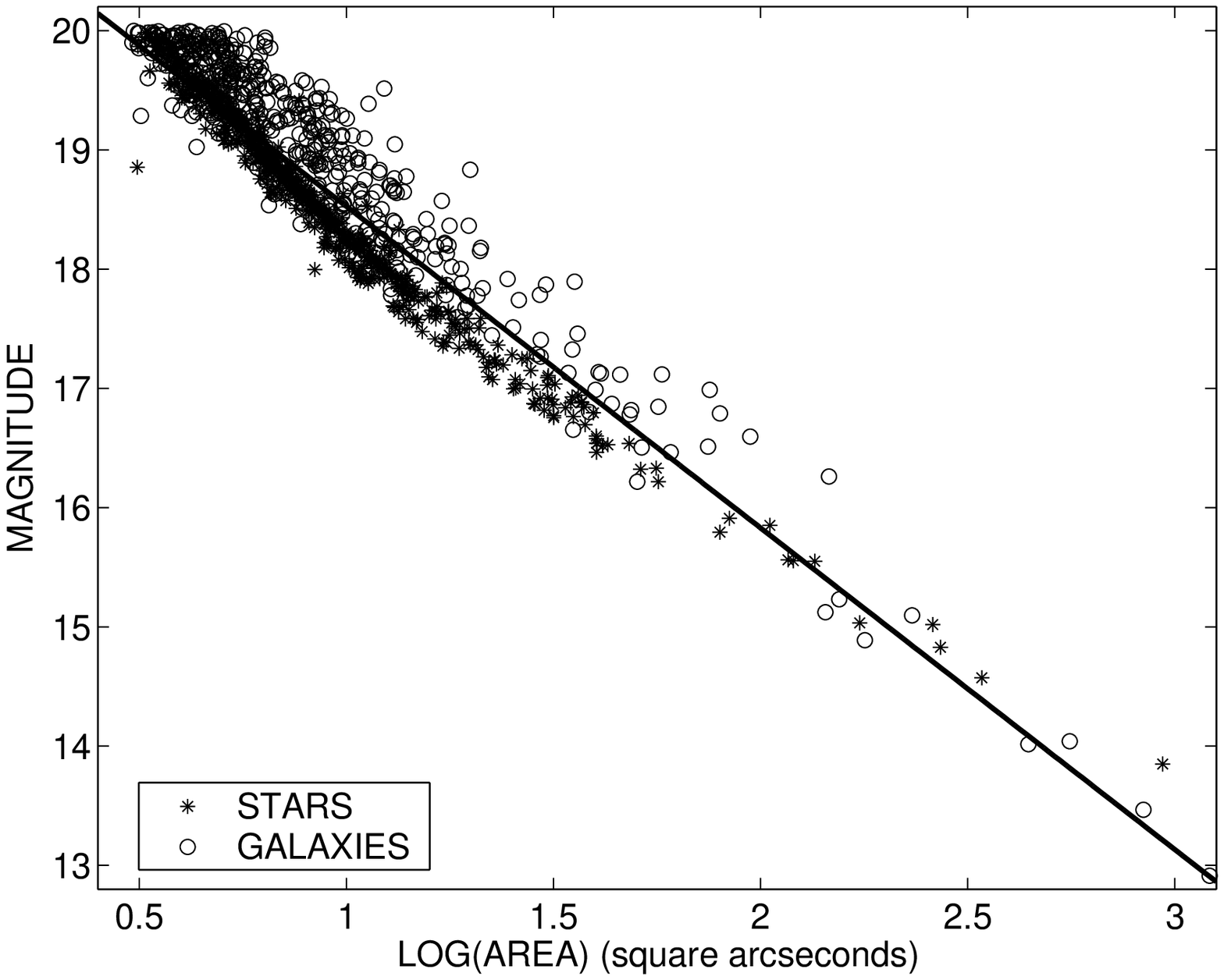}
\includegraphics[width=0.44\textwidth]{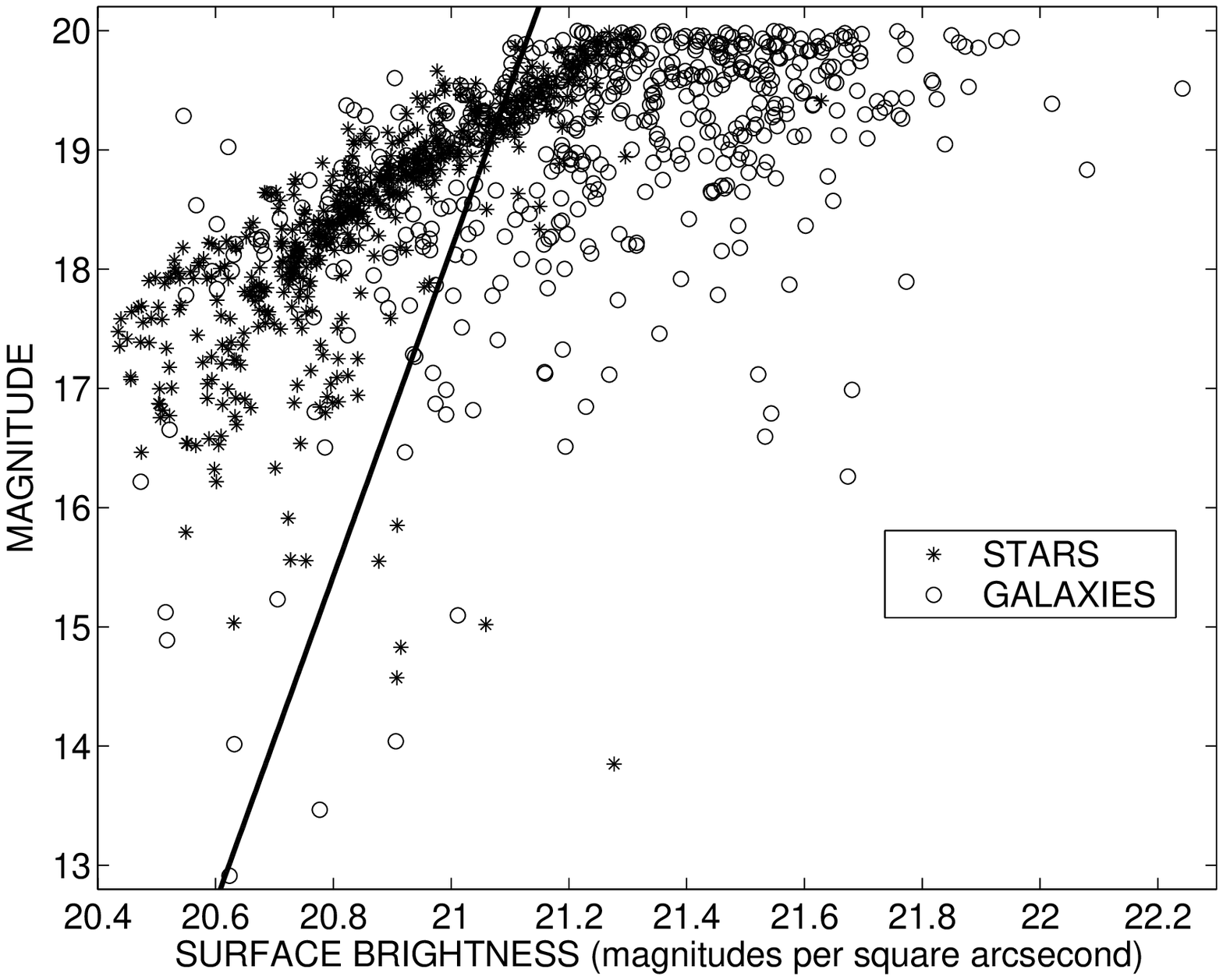}
\includegraphics[width=0.44\textwidth]{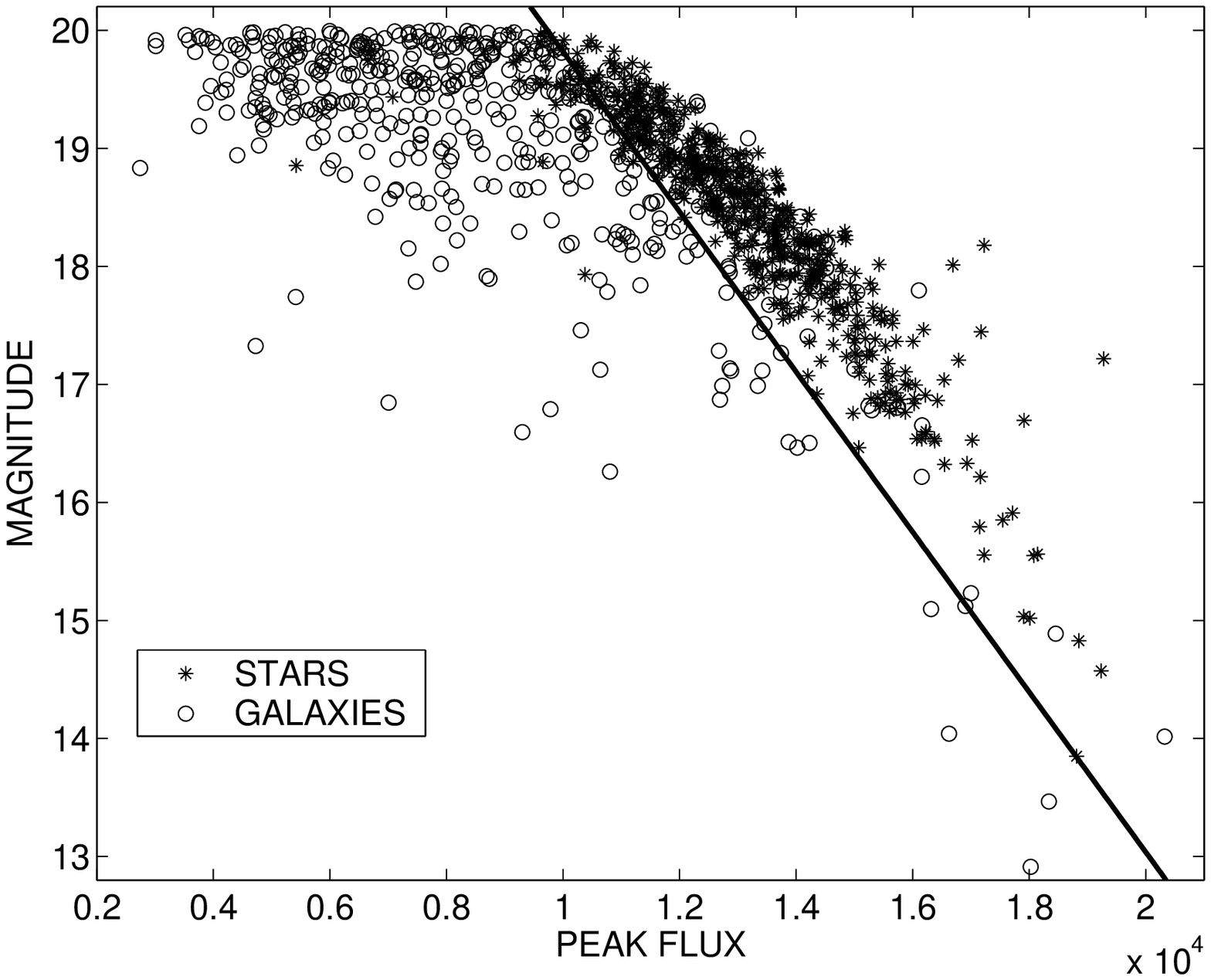}
\caption{\small{Star/galaxy classification plots for photographic plate objects,
between $B$-band  magnitude 20.0 and 13.0.  The top panel shows the
classification using magnitude and aperture area.  The middle panel shows the
classification using magnitude and surface brightness.  Notice that the
classification line has rotated due to low-brightness and high-brightness mixing.
The bottom panel shows the classification using magnitude and peak flux.  The
classification line in each panel has been constructed using an MLP neural
network.  In these three plots, stars brighter than magnitude 16.5 are completely
inseparable from the galaxy population.  These results are summarized in
Table~\ref{tab:photo}.}}
\label{fig:photo1}
\end{figure}

Figure~\ref{fig:photo1} illustrates how magnitude, aperture area, surface
brightness and peak flux can be used to differentiate between stars and galaxies
in photographic plate images.  The magnitude is calculated using the integrated
pixel flux over the aperture, the exposure time, and the photometric zeropoint.
Hence, objects which have saturated to the maximum pixel value will have their
magnitudes under-estimated.  In each plot, the line segment used to separate
stars from galaxies is included --- which has been constructed using an MLP
neural network.  These results are summarized in Table~\ref{tab:photo}.

For the photographic plate images, MM cannot be used to classify stars and
galaxies in the same way that it can for CCD images.  The `peakiness' that MM is
able to detect in the CCD images does not exist in the photographic images, due
to the non-linear response of the plate.

The biggest difficulties for current techniques are differentiating between stars
and galaxies at the high-brightness limit and low brightness limit.  This
high-brightness limitation, best illustrated in Figure~\ref{fig:photo1} (middle
panel), results from the saturation of the stars' profiles, causing them to
become extended like the profile of a galaxy.  The low brightness limitation
results from the resolution of the image, where there just are not enough pixels
to accurately characterize the object.

\begin{figure}[htbp]
\centering
\includegraphics[width=0.44\textwidth]{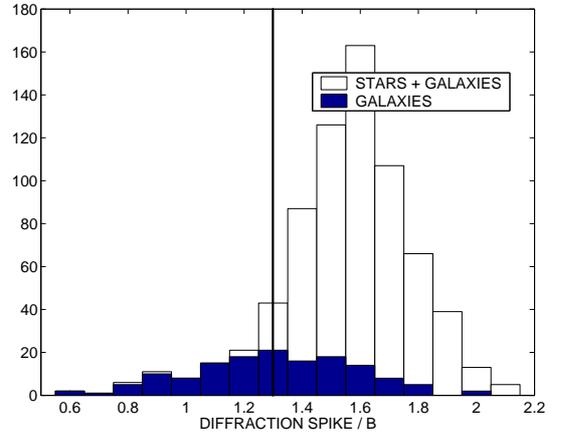}
\caption{\small{Diffraction spike length, calculated using MM, for the bright
stars and galaxies found in the inseparable region of Figure~\ref{fig:photo1}
(middle panel).  The classification line has been constructed using an MLP neural
network.  Galaxies that are unable to be separated from bright stars (using
current techniques alone) can be recovered using the MM diffraction spike tool.
These results are included in Table~\ref{tab:photo}.}}
\label{fig:photo2}
\end{figure}

We use the tools of MM to calculate the sizes of the diffraction spikes, which
become substantial in the stars brighter than magnitude 16.5.  The power of the
structuring element in MM is the ability to match against features within an
image.  We use an increasing series of structuring elements in the shape of a
diffraction spike (plus sign), and the technique of granulometry, to calculate
the diffraction spike length.  This feature presents a method for differentiating
galaxies from bright stars.  Figure~\ref{fig:photo2} illustrates how this has
been achieved, with the objects in this plot taken from the region in
Figure~\ref{fig:photo1} (middle panel) where distinction between stars and
galaxies is difficult, with magnitudes brighter than 17.0 and surface
brightnesses brighter than 20.9 magnitudes per square arcsecond.  The line
segment used to separate stars from galaxies is included --- which has been
constructed using an MLP neural network.  These results are included in
Table~\ref{tab:photo}.

\begin{table}[htbp]
\begin{center}
\caption{Star/Galaxy Differentiation results for photographic plate
images$^a$.}\label{tab:photo}
\begin{tabular}{lrrr}
\hline
Method & Comp. (\%) & Cont. (\%) & Corr. (\%) \\
\hline
Mag--Area  & $22.4 \pm 4.0$ & $1.6 \pm 0.5$ & $83.2 \pm 3.4$ \\
Mag--SB & $22.4 \pm 4.0$ & $1.6 \pm 0.5$ & $83.2 \pm 3.4$ \\
Mag--PF & $21.4 \pm 3.9$ & $1.2 \pm 0.5$ & $83.4 \pm 3.4$ \\
\hline
\end{tabular}
\medskip\\
$^a$ Using 713 high-brightness objects --- 570 stars and 143 galaxies.
\end{center}
\end{table}

\subsection{Results}\label{stargalaxy:results}

For CCD images, $99.3 \pm 3.8$\% of galaxies can be separated from stars using
MM, with $19.4 \pm 7.9$\% of the stars being misclassified, comparing very
favourably with current techniques.  This result, along with the fact that MM
requires only one `free parameter' compared with at least five for techniques 
involving apertures, presents MM as a powerful method for star/galaxy
differentiation in CCD images.

For photographic plate images, conventional techniques fail to differentiate
between stars and galaxies at the high-brightness limit.  A solution to this will
either involve human intervention or a complex algorithm.  MM provides provides a
far less complex solution, involving less `free parameters' than other complex
alternatives.  Using the MM diffraction spike tool allows $51.0 \pm 6.0$\% of the
galaxies that are inseparable in Figure~\ref{fig:photo1} (middle panel), with
magnitudes brighter than 17.0 and surface brightnesses brighter than 20.9
magnitudes per square arcsecond, to be correctly separated from the bright stars.


\section{Galaxy Morphology\\Classification}\label{galaxymorph}

To test the advantage of using MM to classify galaxy morphology we now
exclusively use CCD imaging.  The CCD images are again taken from the data
archive of \citet{sma97}.  It must be noted that peculiar and merging galaxies
are not included in this catalogue, so the ability of classification techniques
to discriminate between such galaxy types is not able to be determined.  For the
images, the techniques of S\'{e}rsic fitting (using the GALFIT software
package\footnote{http://zwicky.as.arizona.edu/$\sim$cyp/work/galfit/galfit.html}),
the `CAS' parameters, and the MM tools, average size and average roughness, are
used to differentiate between pre-classified elliptical (E) and late-type spiral
(Sc-Sd) galaxies.  Before calculating average size and average roughness, an MM 
close operation is performed to average any spiral arm features into a disk
component.  To calculate average size and average roughness, an increasing series
of diamond structuring elements is used for reasons given in
Section~\ref{elements}.

\begin{figure}[htbp]
\centering
\includegraphics[width=0.44\textwidth]{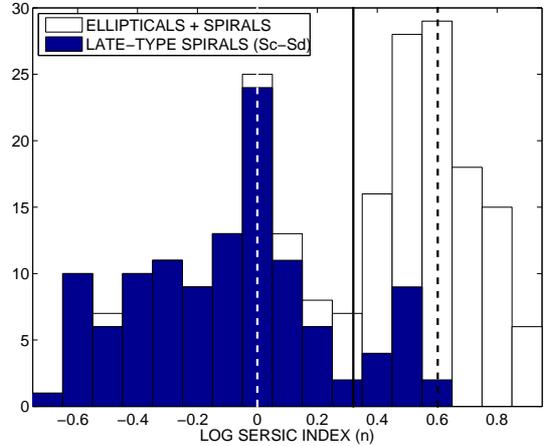}
\caption{\small{Galaxy morphology classification plot for S\'{e}rsic fitting,
using the elliptical (E) and late-type spiral (Sc-Sd) galaxies in the data
archive of \citet{sma97}.  The two dotted lines represent the de Vaucouleurs
(bulge, $n = 4$) and Exponential (disk, $n = 1$) profiles.  The full line
represents the classification line, which has been constructed using an MLP
neural network.  This plot demonstrates the classification performance of
S\'{e}rsic fitting, using the GALFIT software package, in separating elliptical
and late-type spiral galaxies between magnitude 23.5 and 16.0.  These results are
summarized in Table~\ref{tab:gmorph}.}}
\label{fig:gmorph1}
\end{figure}

\begin{figure}[htbp]
\centering
\includegraphics[width=0.44\textwidth]{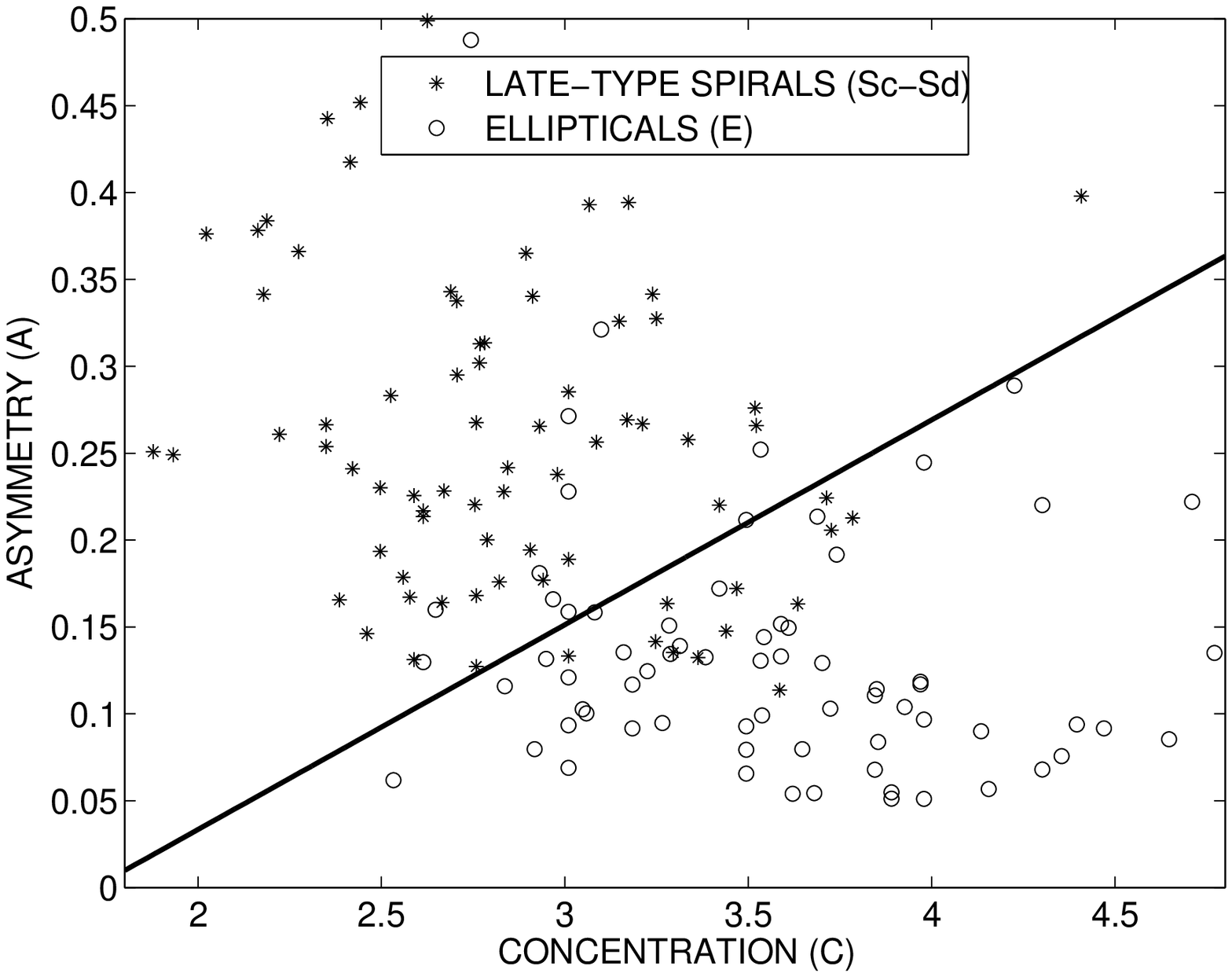}
\includegraphics[width=0.44\textwidth]{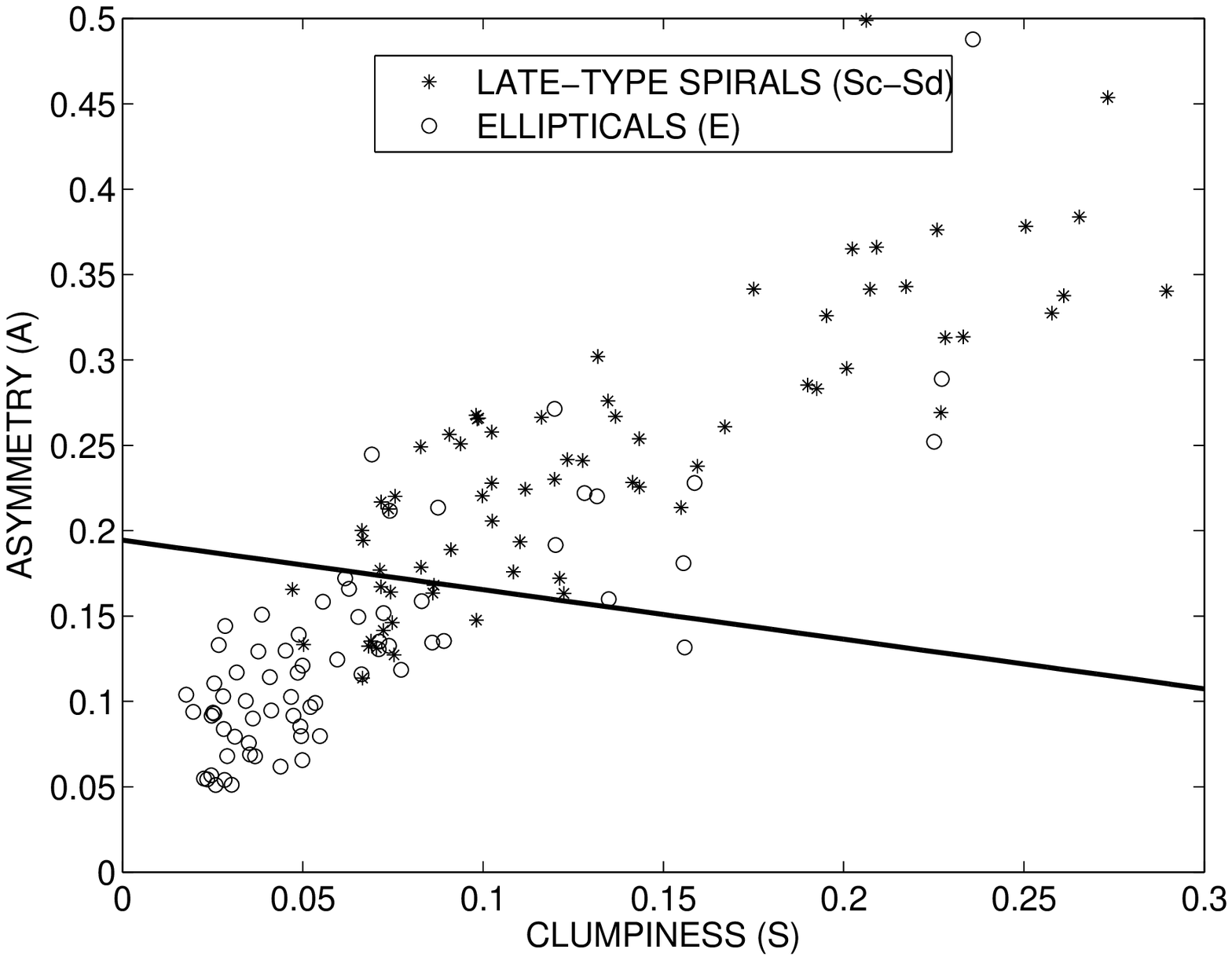}
\includegraphics[width=0.44\textwidth]{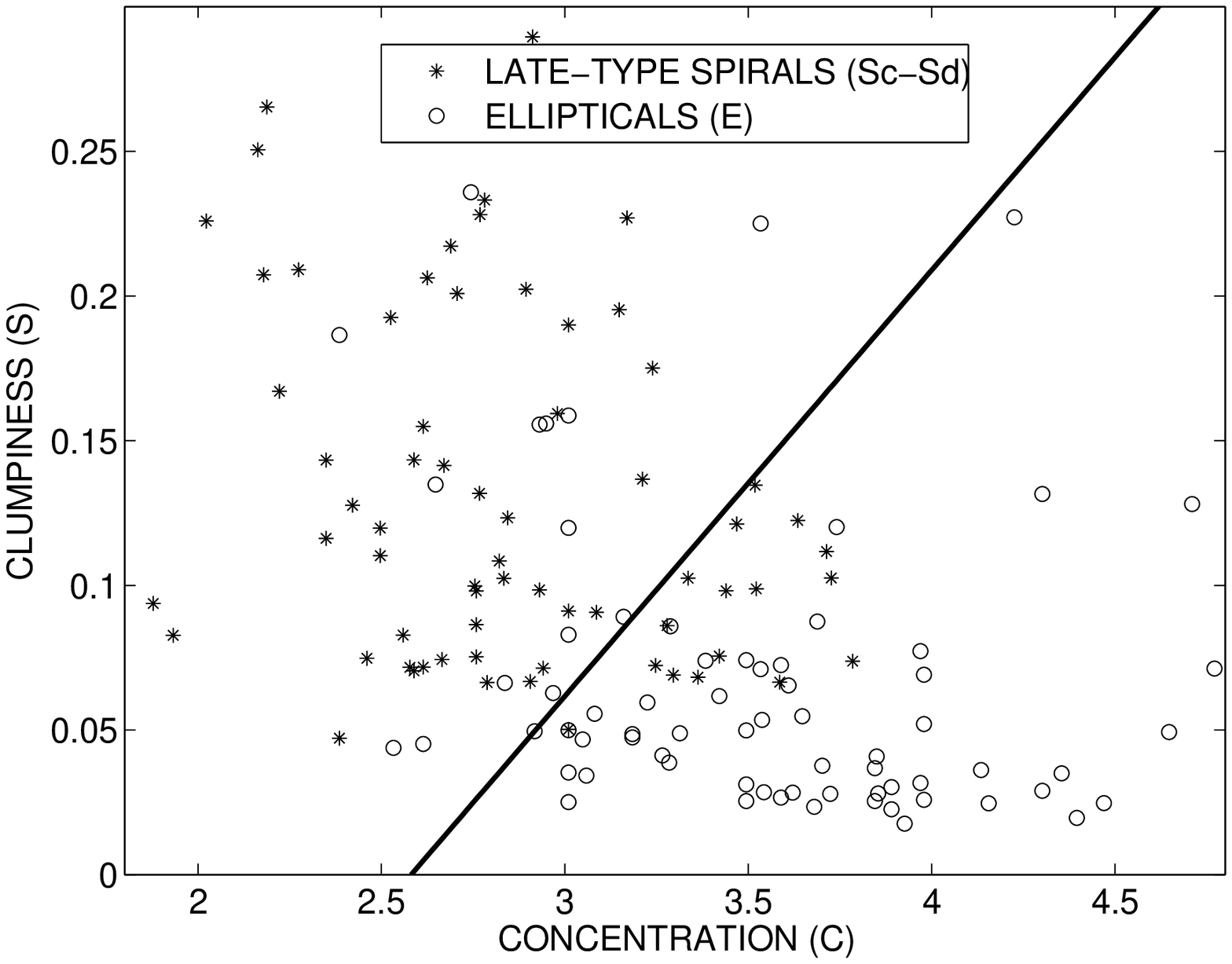}
\caption{\small{Galaxy morphology classification plots for the `CAS' parameters,
using the elliptical (E) and late-type spiral (Sc-Sd) galaxies in the data
archive of \citet{sma97}.  The classification line in each panel has been
constructed using an MLP neural network.  The plots demonstrate the
classification performance of the three `CAS' parameters, in separating
elliptical and late-type spiral galaxies between magnitude 23.5 and 16.0.
These results are included in Table~\ref{tab:gmorph}.}}
\label{fig:gmorph2}
\end{figure}

Figure~\ref{fig:gmorph1} illustrates how the S\'{e}rsic index, $n$, and
Figure~\ref{fig:gmorph2} illustrates how the `CAS' parameters can be used to
classify elliptical and late-type spiral galaxies.  In each plot, the line
segment used to separate elliptical galaxies from late-type spiral galaxies is
included --- which has been constructed using an MLP neural network.  The
galaxies with a core size less than 4 pixels were excluded from the calculation.
We found a severly undersampled core can create difficulties with the clumpiness
and asymmetry parameters, not providing a fair comparison against the other
techniques.  These results are summarized in Table~\ref{tab:gmorph}.

\begin{figure}[htbp]
\centering
\includegraphics[width=0.44\textwidth]{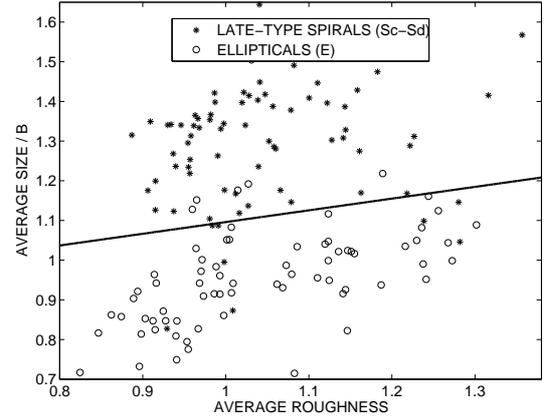}
\caption{\small{Galaxy morphology classification plot for MM, using the
elliptical (E) and late-type spiral (Sc-Sd) galaxies in the data archive of
\citet{sma97}.  The plot shows galaxy morphology classification using two
parameters from MM --- average size and average roughness.  The classification
line has been constructed using an MLP neural network.  The classification
performance of MM is comparable to current techniques, in separating elliptical
and late-type spiral galaxies between magnitude 23.5 and 16.0.  These results are
included in Table~\ref{tab:gmorph}.}}
\label{fig:gmorph3}
\end{figure}

Figure~\ref{fig:gmorph3} shows how MM can be used to classify elliptical and
late-type spiral galaxies.  The line segment used to separate elliptical galaxies
from late-type spiral galaxies is included --- which has been constructed using
an MLP neural network.  These results are included in Table~\ref{tab:gmorph}.

\begin{table}[htbp]
\begin{center}
\caption{Galaxy Morphology Classification results for CCD
images$^a$.}\label{tab:gmorph}
\begin{tabular}{lr}
\hline
Method & Correctness (\%) \\
\hline
S\'{e}rsic Fitting (GALFIT) & $89.8 \pm 6.3$ \\
Asymmetry--Concentration & $84.2 \pm 7.4$ \\
Asymmetry--Clumpiness & $80.9 \pm 7.3$ \\
Clumpiness--Concentration & $79.6 \pm 7.2$ \\
MM & $91.4 \pm 7.8$ \\
\hline
\end{tabular}
\medskip\\
$^a$ Using 152 objects --- 71 elliptical (E) and 81 late-type spiral (Sc-Sd)
galaxies.
\end{center}
\end{table}

\begin{figure}[p]
\centering
\includegraphics[width=0.44\textwidth]{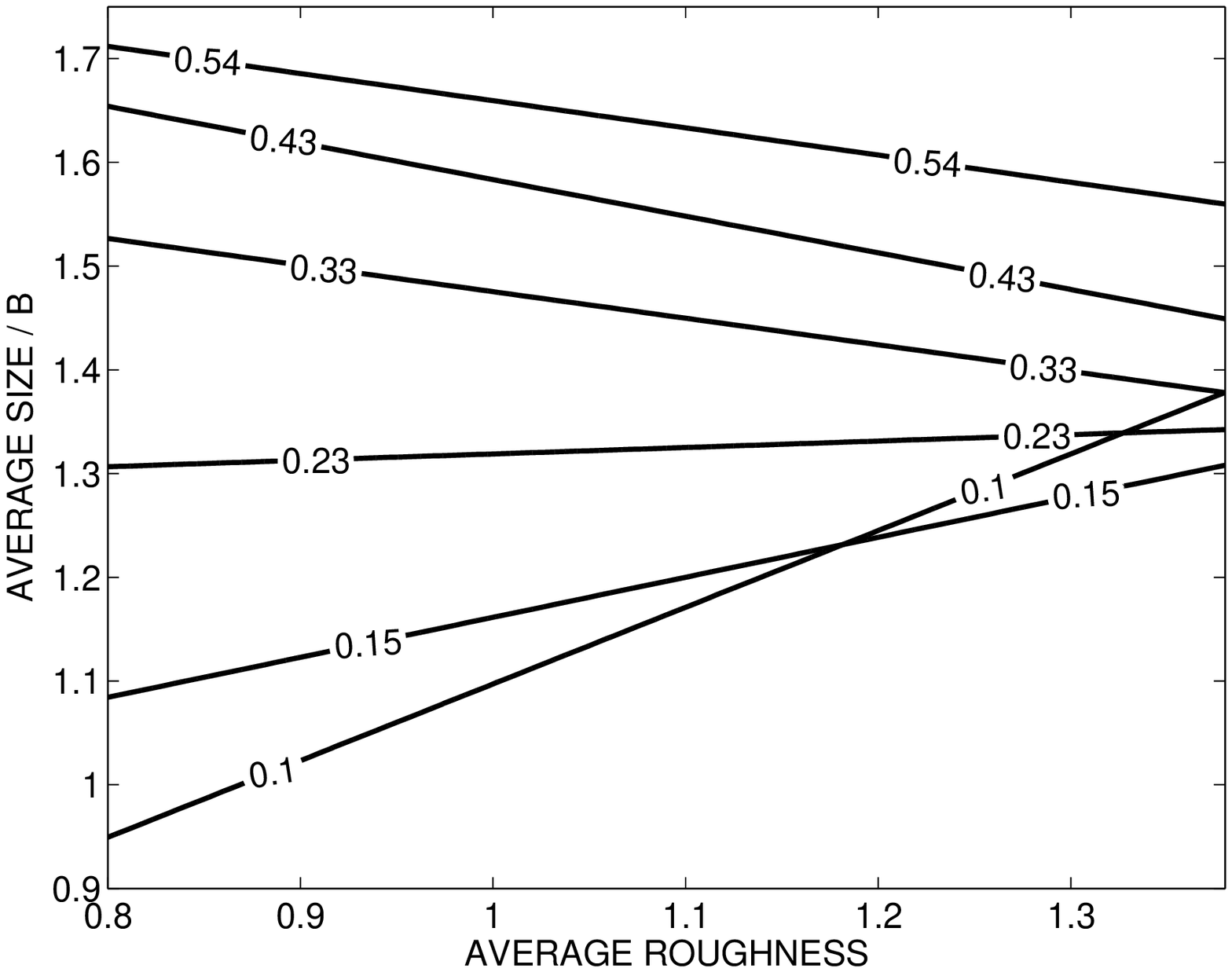}
\includegraphics[width=0.44\textwidth]{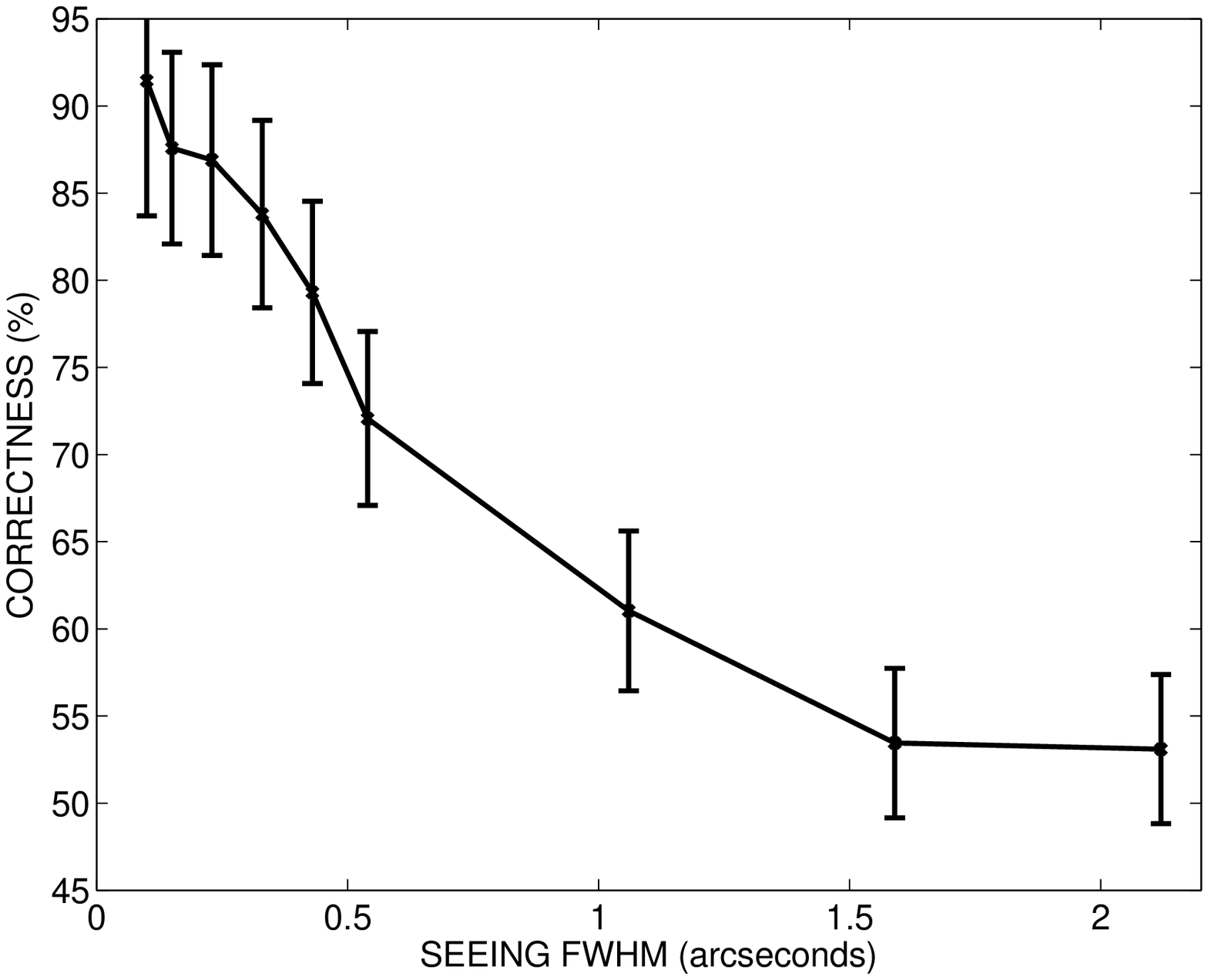}
\caption{\small{Effect of atmospheric smoothing on the MM galaxy morphology
classification.  The top panel shows the change in the classification line
between space-based (FWHM of 0.1 arcsec) to ground-based (FWHM of 0.54 arcsec)
seeing conditions.  The bottom panel shows the reduction in the correctness of
the classification between space-based (FWHM of 0.1 arcsec) to ground-based
(FWHM of 2.12 arcsec) seeing conditions.  The performance is acceptable (at least
80\%) for seeing better than 0.4 arcsec.}}
\label{fig:gmorph4}
\end{figure}

To test the effect of atmospheric smoothing (or seeing) on the performance of the
MM galaxy morphology classification, the {\it HST} images are artificially
blurred using the {\textsc IRAF} Gaussian task from space-based (FWHM of 0.1
arcsec) to ground-based (FWHM of 2.12 arcsec) seeing conditions.  The left panel
of Figure~\ref{fig:gmorph4} shows how the classification line is altered as the
seeing increases, shifting toward higher average size values as the galaxy
profiles extend due to the blurring.  The right panel of Figure~\ref{fig:gmorph4}
shows how the classification performance of MM is reduced as the seeing increases.

\subsection{Results}\label{galaxymorph:results}

For CCD images,  $91.4 \pm 7.8$\% of elliptical and late-type spiral galaxies can
be correctly classified using MM, comparing very favourably with current
techniques.  These results present MM as a powerful alternative for galaxy
morphology classification in CCD images.

Atmospheric seeing effects the MM galaxy morphology classification by shifting
the classification line toward higher average size values and reducing the
correctness of the classification.  The performance of the classification is
acceptable (at least 80\%) for seeing FWHM equal to or better than 0.4
arcseconds.  For a redshift of $0.465 \pm 0.095$ ($D_A = 1193.1_{-152.1}^{+122.4}$
Mpc; for $H_0 = 71$, $\Omega_M = 0.3$, $\Omega_\Lambda = 0.7$) this corresponds
to a physical size of $2.19_{-0.29}^{+0.24}$ kpc.  This size is of the same order
as the central bulge of each galaxy and, therefore, the Gaussian seeing profile
will become a significant element of the galaxy profiles.  The difference between 
the profiles of elliptical and spiral galaxies is reduced with seeing sizes even
larger, preventing MM from being able to distinguish between the two classes.
From this we can write the following seeing-distance relationship --- MM will
classify galaxy morphology at a performance of 80\% or greater if:
\begin{equation}
   s \times d < 477.24_{-60.84}^{+48.96}
\end{equation}
where $s$ is the seeing FWHM (in arc seconds) and $d$ is the cosmological angular
size distance (in Mpc).  The consequence of this is that, for a ground-based
telescope with a 1.0 arcsecond seeing FWHM, the MM classification is acceptable
to a maximum distance of $477.24_{-60.84}^{+48.96}$ Mpc ($z =  0.13 \pm 0.02$).

The difficulty remains to find what physical processes the average size and
average roughness will correlate with.  For example, using the `CAS' framework,
\citet{con03} use some quantitative reasoning to show that the concentration index
correlates with gross form (bulge to disk ratio), asymmetry correlates with recent
merger/interaction activity, and clumpiness correlates with star formation (using
the H$\alpha$ equivalent widths).  We are currently investigating the relationship
between average size and average roughness with respect to star formation rate in
galaxies.  This will provide correlation with both internal and external galaxy
mechanisms, and will be presented in a future publication.


\section{Summary}\label{summary}

We have demonstrated that, for CCD images, $99.3 \pm 3.8$\% of the galaxy
population is separated from the stellar population using MM, with
$19.4 \pm 7.9$\% of the stars being misclassified, comparing very well with
current techniques involving aperture magnitudes.  We have also demostrated that,
for photographic plate images, the MM diffraction spike tool allows
$51.0 \pm 6.0$\% of the bright galaxies, that are inseparable from the bright
stars in current techniques, to be correctly classified, with only
$1.4 \pm 0.5$\% of the high-brightness stars contaminating the population.
Lastly, we have demonstrated that elliptical (E) and late-type spiral (Sc-Sd)
galaxies can be classified using MM at an accuracy of $91.4 \pm 7.8$\%, which
compared well with the techniques of S{\'e}rsic fitting and the `CAS' parameters.

What we have presented here is an automated method, alternative to current 
techniques, for both star/galaxy differentiation and galaxy morphology
classification.  It is a method involving less `free parameters' than current
techniques, especially automated machine learning algorithms, and doesn't require
a machine learning algorithm for the training of the classifier.  The
classification tools developed in this work could, however, also be used as an
additional layer of parameter space for any current machine learning approaches.

We intend to use this method to study the relationship between galaxy morphology
and star formation rate, and how each vary within the supercluster environment.
This environment allows us to probe the complete range of densities at which
galaxies are affected by star formation suppression and morphology alteration
mechanisms.  We shall study how these galaxy properties vary between dense
cluster cores, inter-cluster filaments, and the low-density `field' regions.

\section*{Acknowledgments}

We thank the anonymous referees for their careful reading and thoughtful comments
which have improved the quality of this work.  We also kindly thank Marianne
Doyle and David Rhode for providing us with the SuperCOSMOS images, along with the
SExtractor segmentation and magnitude calibration for each field.  We also thank
useful conversations with various members of the Astronomy Group at the University
of Queensland.  J.A.M. acknowledges financial support from an EPSA University of
Queensland Joint Research Scholarship.  K.A.P. acknowledges financial support
from an EPSA University of Queensland Research Fellowship.

\section*{Appendix A:\\`CAS' Volume}\label{app:cas}

Here, we give just a brief summary of each parameter, but a more detailed description
of the operation of these can be found in the work of \citet{con03}.

\paragraph{Concentration of Light}
\ \\
The concentration index, $C$, is defined as \citep{ber00} the ratio of the 80\% to
20\% curve of growth radii, normalized by a logarithm:
\begin{equation}
   C = 5 \times \log(\frac{r_{80\%}}{r_{20\%}})
\end{equation}
where $r_{x\%}$ represents the size of the aperture which contains $x\%$ of the total
light flux of the galaxy (in some implementations the 75\% and 25\% curve of growth
radii are used instead).

\paragraph{Asymmetry}
\ \\
The asymmetry index, $A$, is defined as \citep{con00} the volume of the image of a
galaxy which has been rotated $180^{\circ}$ around its center and then subtracted
from its pre-rotated image, then normalized to the original image volume:
\begin{equation}
   A = \frac{\sum \vert I - R \vert}{\sum \vert I \vert}
\end{equation}
where $I$ is the original image and $R$ is the rotated image.  \citet{con03} first
reduce the effective resolution of the image, $I$, using a filter of size
$\frac{1}{6} \times r(\eta = 0.2)$, to have asymmetry only sensitive to large-scale
stellar distributions.

\paragraph{High Spatial Frequency}
\ \\
The clumpiness parameter, $S$, is defined as \citep{con03} the amount of light
contained in high spatial frequency structures to the total amount of light in the
galaxy.  Computationally, this is defined as the volume of the image of a galaxy which
has been blurred (smoothed) using a filter and then subtracted from its pre-smoothed
image, then normalized to the original image volume:
\begin{equation}
   S = \frac{\sum (I - B)}{\sum I}
\end{equation}
where $I$ is the original image and $B$ is the blurred (smoothed) image.
\citet{con03} use a smoothing filter of size
$\frac{1}{5} \times 1.5 \times r(\eta = 0.2)$ and exclude the region inside
$\frac{1}{20} \times 1.5 \times r(\eta = 0.2)$, as its high-frequency power is
unrelated to the stellar light distribution.  They also force any negative pixel
values in the difference image, $I - B$, to zero before computing the clumpiness.


\begin{thebibliography}{}
\bibitem[Abraham et al.(1994)]{abr94} Abraham, R.~G., Valdes, F., Yee, H.~K.~C.,
\& van den Bergh, S. 1994, ApJ, 432, 75
\bibitem[Andredakis \& Sanders(1994)]{and94} Andredakis, Y.~C., \& Sanders, R.~H.
1994, MNRAS, 267, 283
\bibitem[Andreon et al.(2000)]{and00} Andreon, S., Gargiulo, G., Longo, G.,
Tangliaferri, R., \& Capuano, N. 2000, MNRAS, 319, 700
\bibitem[Appleton, Siqueira, \& Basart(1993)Appleton et al.]{app93} Appleton, P.~N.,
Siqueira, P.~R., \& Basart, J.~P. 1993, AJ, 106, 1664
\bibitem[Bershady, Jangren, \& Conselice(2000)Bershady et al.]{ber00} Bershady, M.~A.,
Jangren, A., \& Conselice, C.~J. 2000, AJ, 119, 2645
\bibitem[Bertin \& Arnouts(1996)]{ber96} Bertin, E., \& Arnouts, S. 1996,
A\&AS, 117, 393
\bibitem[Candeas, Braga Neto, \& Carvalho Filho(1996)Candeas et al.]{can96}
Candeas, A.~J., Braga Neto, U., \& Carvalho Filho, E. 1996, Anais do IX SIBGRAPI, 235
\bibitem[Candeas, Braga Neto, \& Carvalho Filho(1997)Candeas et al.]{can97}
Candeas, A.~J., Braga Neto, U., \& Carvalho Filho, E. 1997, J. Braz. Comp. Soc., 3
\bibitem[Conselice, Bershady, \& Jangren(2000)Conselice et al.]{con00}
Conselice, C.~J., Bershady, M.~A., \& Jangren, A. 2000, ApJ, 529, 886
\bibitem[Conselice(2003)]{con03} Conselice, C.~J. 2003, ApJS, 147, 1
\bibitem[de Vaucouleurs(1948)]{vau48} de Vaucouleurs, G. 1948, AnAp, 11, 247
\bibitem[de Vaucouleurs(1959)]{vau59} de Vaucouleurs, G. 1959,
Handb. der Physik, 53, 275
\bibitem[Doyle et al.(2005)]{doy05} Doyle, M.~T., et al. 2005, MNRAS, 361, 34
\bibitem[Drinkwater \& Schmidt(1996)]{dri96} Drinkwater, M.~J. \& Schmidt, R.~W. 1996,
PASA, 13, 127
\bibitem[Graham \& Driver(2005)]{gra05} Graham, A.~W., \& Driver, S.~P. 2005,
PASA, 22, 118
\bibitem[Hambly et al.(2001a)]{ham01a} Hambly, N.~C., et al. 2001a, MNRAS, 326, 1279
\bibitem[Hambly, Irwin, \& MacGillivray(2001b)Hambly et al.]{ham01b} Hambly, N.~C.,
Irwin, M.~J., \& MacGillivray, H.~T. 2001b, MNRAS, 326, 1295
\bibitem[Hambly et al.(2001c)]{ham01c} Hambly, N.~C., Davenhall, A.~C., Irwin, M.~J.,
\& MacGillivray, H.~T. 2001c, MNRAS, 326, 1315
\bibitem[Harmon \& Mamon(1993)]{har93} Harmon, R., \& Mamon, G. 1993,
in ASP Conf. Ser. 43: Sky Surveys: Protostars to Protogalaxies,
ed. B.~T. Soifer (San Francisco: ASP), 15
\bibitem[Haykin(1998)]{hay98} Haykin, S. 1998,
Neural Networks - A Comprehensive Foundation, 2nd ed. (Englewood Cliffs: Prentice-Hall)
\bibitem[He(1996)]{he96} He, L.~X. 1996, Ph.D. Thesis, Iowa State University
\bibitem[Hubble(1926)]{hub26} Hubble, E. 1926, ApJ, 64, 321
\bibitem[Hubble(1936)]{hub36} Hubble, E. 1936,
The Realm of the Nebulae (New Haven: Yale University Press)
\bibitem[Heijmans(1992)]{hei92} Heijmans, H.~J.~A.~M. 1992,
Nieuw Archief voor Wiskunde, Vierde Serie, 10, 237
\bibitem[Heijmans(1994)]{hei94} Heijmans, H.~J.~A.~M. 1994,
in Shape in Picture: Mathematical Description of Shape in Grey-level Images, 147
\bibitem[Heijmans(1995)]{hei95} Heijmans, H.~J.~A.~M. 1995, in SIAM Review, 37, 1
\bibitem[Jones et al.(1991)]{jon91} Jones, L.~R., Fong, R., Shanks, T., Ellis, R.~S.,
\& Peterson, B.~A. 1991, MNRAS, 249, 481
\bibitem[Kibblewhite et al.(1984)]{kib84} Kibblewhite, E.~J., Bridgeland, M.~T.,
Bunclark, P., \& Irwin, M.~J. 1984,
in NASA Conf. Publ. 2317: Astronomical Microdensitometry Conference,
ed. D.~A. Klinglesmith (Washington, D.C.: NASA Scientific and Technical Information
Branch), 277
\bibitem[Kron(1980)]{kro80} Kron, R.~G. 1980, ApJS, 43, 305
\bibitem[Lea \& Keller(1989)]{lea89} Lea, S.~M., \& Kellar, L.~A. 1989, AJ, 97, 1238
\bibitem[Lef\`{e}vre et al.(1986)]{lef86} Lef\`{e}vre, O., Bijaoui, A., Mathez, G.,
Picat, J.~P., \& Lel\i`{e}vre, G. 1986, A\&A, 154, 92
\bibitem[MacArthur, Courteau, \& Holtzman(2003)MacArthur et al.]{mac03}
MacArthur, L.~A., Courteau, S., \& Holtzman, J.~A. 2003, ApJ, 582, 689
\bibitem[Maddox et al.(1990)]{mad90} Maddox, S., Sutherland, W., Efstathiou, G.,
\& Loveday, J. 1990, MNRAS, 243, 692
\bibitem[M\"ah\"onen \& Frantti(2000)]{mah00} M\"ah\"onen, P., \& Frantti, T. 2000,
ApJ, 541, 261
\bibitem[Matheron(1975)]{mat75} Matheron, G. 1975,
Random Sets and Integral Geometry (New York: John Wiley and Sons)
\bibitem[Maragos(1989)]{mar89} Maragos, P. 1989,
IEEE Trans. Pattern Anal. Mach. Intell., 11, 701
\bibitem[Miller \& Coe(1996)]{mil96} Miller, A.S., \& Coe, M.~J. 1996, MNRAS, 279, 293
\bibitem[Odewahn et al.(1992)]{ode92} Odewahn, S., Stockwell, E., Pennington, R.~M.,
Humphreys, R., \& Zumach, W. 1992, AJ, 103, 318
\bibitem[Odewahn et al.(1993)]{ode93} Odewahn, S., Humphreys, R.~M., Aldering, G.,
\& Thurmes, P. 1993, PASP, 105, 1354
\bibitem[Philip et al.(2002)]{phi02} Philip, N.~S., Wadadekar, Y., Kembhavi, A.,
\& Joseph, K.~B. 2002, A\&A, 385, 1119
\bibitem[Pimbblet et al.(2001)]{pim01} Pimbblet, K.~A., Smail, I., Edge, A.~C.,
Couch, W.~J., O'Hely, E., \& Zabludoff, A.~I. 2001, MNRAS, 327, 588
\bibitem[Peng et al.(2002)]{pen02} Peng, C.~Y., Ho, L.~C., Impey, C.~D.,
\& Rix, H-W. 2002, AJ, 124, 266
\bibitem[Reid \& Gilmore(1982)]{rei82} Reid, N. \& Gilmore, G. 1982, MNRAS, 201, 73
\bibitem[Reid et al.(1996)]{rei96} Reid, I.~N., Yan, L., Majewski, S., Thompson, I.,
\& Smail, I. 1996, AJ, 112, 1472
\bibitem[Sandage (1961)]{san61} Sandage, A. 1961,
The Hubble Atlas of Galaxies (Washington, D.C.: Carnegie Institution of Washington)
\bibitem[Shaver(1987)]{sha87} Shaver, P.~A. 1997, Nature, 326, 773
\bibitem[Sebok(1979)]{seb79} Sebok, W. 1979, AJ, 84, 1526
\bibitem[Serra(1982)]{ser82} Serra, J. 1982,
Image Analysis and Mathematical Morphology (London: Academic Press)
\bibitem[S\'{e}rsic(1963)]{ser63} S\'{e}rsic, J.~L. 1963, BAAA, 6, 41
\bibitem[S\'{e}rsic(1968)]{ser68} S\'{e}rsic, J.~L. 1968,
Atlas de Galaxias Australes (Cordoba: Observatorio Astronomico)
\bibitem[Smail et al.(1997)]{sma97} Smail, I., Dressler, A., Couch, W.~J.,
Ellis, R.~S., Oemler, A. (Jr), Butcher, H., \& Sharples, R.~M. 1997, ApJS, 110, 213
\bibitem[Ueda(1999)]{ued99} Ueda, H. 1999, PASJ, 51, 435
\bibitem[van der Bergh(1960)]{ber60} van der Bergh, S. 1960, ApJ, 131, 215
\bibitem[Weir et al.(1995)]{wei95} Weir, N., Fayyad, U.~M., \& Djorgovski, S. 1995,
AJ, 109, 6
\end{thebibliography}
\end{document}